\documentclass[twocolumn,prl,aps,superscriptaddress,showpacs]{revtex4}

\usepackage{bm}
\usepackage{graphicx}
\usepackage{color}

\newcommand{\urs}{URu$_2$Si$_2$}

\begin{document}

\title{Suppression of hidden order in URu$_2$Si$_2$ under pressure and restoration in magnetic field}

\author{E Hassinger, D Aoki, F Bourdarot, G Knebel, V Taufour, S Raymond, A Villaume and J Flouquet}

\affiliation{CEA Grenoble, 17 rue de Martyrs, 38000 Grenoble, France}


\begin{abstract}
We describe here recent inelastic neutron scattering experiments on the heavy fermion compound \urs realized in order to clarify the nature of the hidden order (HO) phase which occurs below $T_0=17.5$\,K at ambient pressure.
The choice was to measure at a given pressure $P$ where the system will go, by lowering the temperature, successively from paramagnetic (PM) to HO and then to antiferromagnetic phase (AF). Furthermore, in order to verify the selection of the pressure, a macroscopic detection of the phase transitions was also achieved in situ via its thermal expansion response detected by a strain gauge glued on the crystal.
Just above $P_x=0.5$\,GPa, where the ground state switches from HO to AF, the $Q_0=(1, 0, 0)$ excitation disappears while the excitation at the incommensurate wavevector $Q_1=(1.4, 0, 0)$ remains. Thus, the $Q_0=(1, 0, 0)$ excitation is intrinsic only in the HO phase. This result is reinforced by studies where now pressure and magnetic field $H$ can be used as tuning variable. Above $P_x$, the AF phase at low temperature is destroyed by a magnetic field larger than $H_{AF}$ (collapse of the AF $Q_0=(1, 0, 0)$ Bragg reflection). The field reentrance of the HO phase is demonstrated by the reappearance of its characteristic $Q_0=(1, 0, 0)$ excitation. The recovery of a PM phase will only be achieved far above $H_{AF}$ at $H_M \approx 35$\,T. To determine the P-H-T phase diagram of URu$_2$Si$_2$, macroscopic measurements of the thermal expansion were realized with a strain gauge. The reentrant magnetic field increases strongly with pressure.
Finally, to investigate the interplay between superconductivity (SC) and spin dynamics, new inelastic neutron scattering experiments are reported down to 0.4\,K, far below the superconducting critical temperature $T_{SC} \approx 1.3$\,K as measured on our crystal by diamagnetic shielding.
\end{abstract}
\maketitle
\section{Introduction}
URu$_2$Si$_2$ is a heavy fermion compound with a body centered tetragonal crystal structure in its paramagnetic phase. At ambient pressure, on lowering the temperature below $T_0=17.5$\,K a transition occurs into a hidden order phase which cannot be a conventional magnetic phase. Over the last decade, the picture was that of small moment antiferromagnetism (as all neutron scattering experiments point out a tiny ordered moment $M_0 \approx 0.02 \mu_B$ with the propagation vector $Q_{AF} = (0, 0, 1)$ \cite{Broholm1987}) associated with an electronic instability corresponding to a loss of electronic carriers at $T_0$ \cite{Maple1986,Schoenes1987,Behnia2005}. The departure from the conventional AF picture is clearly demonstrated by the fact that the entropy loss at $T_0$, $\Delta S \approx 0.2R\ln(2)$, is too high to be accounted for exclusively by the formation of localized tiny ordered moments. NMR and $\mu$SR spectroscopy rule out an intrinsic origin of the tiny sublattice magnetization \cite{Matsuda2003, Amato2004}; the small moment seems generated by lattice defaults. This high sensitivity to lattice imperfections is clearly related with the observation that at a rather weak pressure $P_x \approx 0.5$\,GPa \cite{Motoyama2003, Bourdarot2004a} the ground state switches from HO to AF ground state with a substantional sublattice magnetization $M_0 = 0.4  \mu_B$ at $T \rightarrow 0$ with the same $Q_{AF}$. Here we tune from HO to AF under pressure to clarify the main properties of each phase and in particular noting their inelastic neutron scattering spectra.\\
\indent At the opposite to the elastic AF neutron scattering response at ambient pressure, the inelastic neutron scattering signals are very robust \cite{Broholm1987}: excellent agreement exists between results. Below $T_0$, inelastic neutron scattering experiments indicate two sharp and intense excitations, one with a symmetric shape in the energy spectrum at the incommensurate position $Q_1 = (1.4, 0, 0)$ with a gap $E_1 \approx 4.8$ meV and another one with an asymmetric shape at the wavevector $Q_0$ = (1, 0, 0) with a gap $E_0 \approx 2$\,meV. Note that the wavevector $Q_0 = (1, 0, 0)$ is in another Brillouin zone equivalent to the ordering wavevector $Q_{AF} = (0, 0, 1)$. On warming above $T_0$, both excitations respond differently. The signal at $Q_0$ is quasielastic and strongly damped, whereas at $Q_1$ it remains inelastic but so strongly damped that the spectrum is not fully gapped any more. The $T$-evolution of the excitation at $Q_1$ and equivalent positions can explain the specific heat anomaly at $T_0$ \cite{Wiebe2007}. It is worthwhile to recognize that if no phase transition occured at $T_0$, \urs would be a classical intermediate valence compound; it is the feedback between spin dynamics and band structure reconstruction which leads to the fact that below $T_0$ sharp excitations at $Q_0$ and $Q_1$ emerge \cite{Hassinger2008a}.

\section{Results and discussion}
\begin{figure}[h!]
\begin{minipage}{17pc}
\includegraphics[width=17pc]{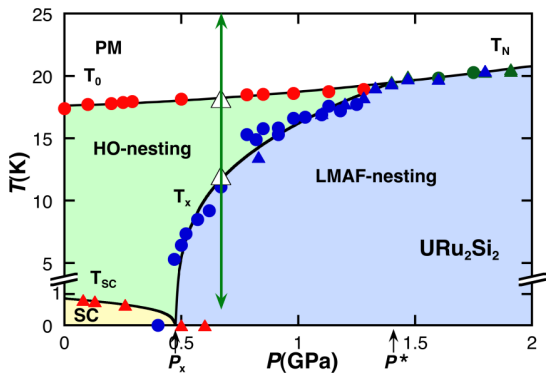}
\end{minipage}\hspace{4pc}%
\begin{minipage}{17pc}
\caption{\label{Fig1}Pressure-temperature phase diagram of URu$_2$Si$_2$ determined by specific heat and resistivity measurements \cite{Hassinger2008}. The green vertical line indicates the pressure where the excitations were measured.}
\end{minipage} 
\end{figure}
\begin{figure}[h]
\includegraphics[width=16pc]{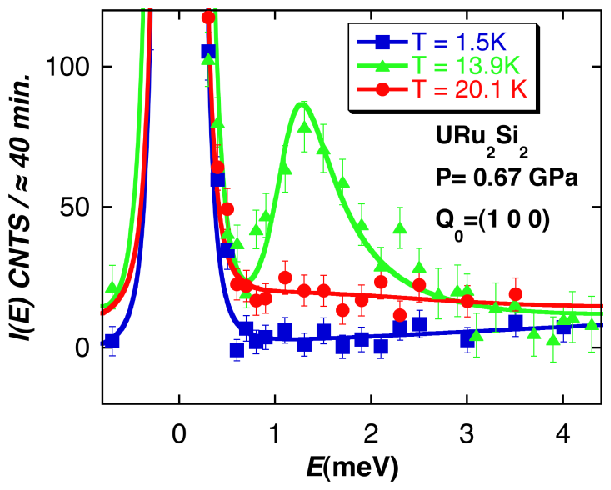}
\includegraphics[width=16pc]{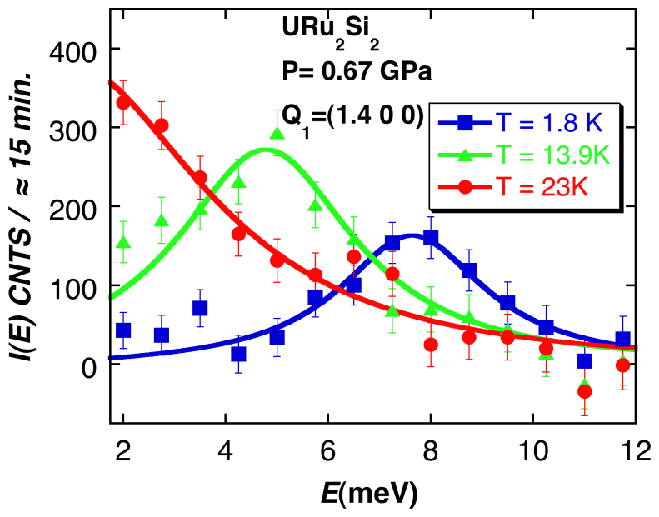}
\caption{\label{Fig2}On the upper panel are shown the energy spectra at the wave vector $Q_0 = (1, 0, 0)$. The excitation present at 13.9\,K in the HO state (green triangles) disappears completely at low temperature in the AF phase (blue squares). On the lower panel the excitation at $Q_1 = (1.4, 0, 0)$. The lines are guides to the eye.\cite{Villaume2008}}
\end{figure}

Figure \ref{Fig1} 
shows the ($T$,$P$) phase diagram of \urs as recently determined by resistivity and microcalorimetry experiments \cite{Hassinger2008}. Special attention was given in the determination of the three lines: $T_0$($P$) of the transition from PM to HO, $T_x$($P$) from HO to AF and $T_N$($P$) from PM to AF. In agreement with thermal expansion and elastic neutron scattering experiments \cite{Motoyama2003,Bourdarot2004a,Amitsuka2007}, $P_x \approx 0.5$\,GPa at $T\rightarrow 0$, while the three lines seem to meet at the critical pressure $P_c \approx 1.3$\,GPa with $T_c \approx 18.5$\,K. The $P$-evolution of the two excitations were reported nine years ago \cite{Amitsuka2000}; four different pressures were measured on crystals where $P_x$ was reported near 1.5\,GPa. Both excitations exist at 0.87\,GPa but disappear simultaneously at 1.86\,GPa. Taking into account progress in the determination of the ($P$,$T$) phase diagram and also our recent ability to measure simultaneously neutron scattering and thermal expansion via a strain gauge in the same pressure cell, we decided to study at a selected pressure $P$ between $P_x$ and $P_c$ (vertical green line in figure \ref{Fig1}) the evolution of the inelastic neutron scattering spectrum in the three phases PM, HO and AF for both $Q_0$ and $Q_1$ using the high performance of the respective triple axis spectrometers IN12 and IN22 at the Institut Laue Langevin (ILL) \cite{Villaume2008}. The figure \ref{Fig2} shows the results for the two wavevectors. Clearly, the collective excitation at $Q_0 = (1, 0, 0)$ is characteristic 
of the HO phase as it disappears below $T_x \approx 13$\,K while the signal at $Q_1 = (1.4, 0, 0)$ remains but shifts in energy. Considering the fact that the $Q_0=(1, 0, 0)$ excitation is characteristic of strong longitudinal fluctuations at the wavevector $Q_{AF}$ and that no drastic difference is detected in the carrier number between both sides of $P_x$ \cite{Hassinger2008}, the proposal is that the Fermi surface reconstruction at $T_0$ and $T_N$ are identical. This statement is supported by our recent study of Shubnikov-de Haas oscillations through $P_x$ [Hassinger {\it et al.} to be published], where it was observed that the three observed frequencies ($\alpha, \beta, \gamma$) are $P$ invariant. In the AF phase the two uranium sites are inequivalent and the symmetry changes. In excellent agreement with band structure calculations \cite{Elgazzar2009,Ohkuni1999,Yamagami2000}, a drop of carrier density is associated with this change. In the HO phase, our measurements strongly suggest that $Q_{AF}$ will also be the wavevector of the HO state. Recently it was stressed out from band structure calculations, that the strong longitudinal fluctuations at $Q_{AF}$ in the HO phase preserve a Fermi surface reconstruction similar to the AF state \cite{Elgazzar2009}.\\
\indent A recent complementary experiment was to continue the inelastic neutron scattering studies with the addition of a magnetic field \cite{Aoki2009}. The experiments were performed on the IN14 spectrometer at the ILL and
\begin{figure}[h]
\includegraphics[width=19pc]{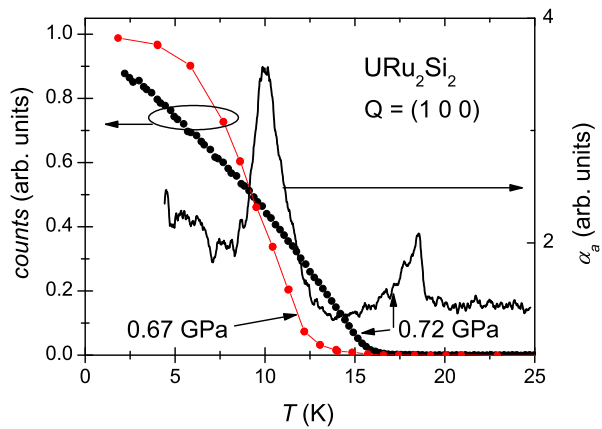}\hspace{4pc}%
\caption{\label{n+t}Scattering intensity at the magnetic Bragg peak position as a function of temperature at a fixed pressure of 0.72\,GPa in zero field (black dots, left scale) compared to the same signal in reference \cite{Villaume2008}(red dots\,+\,line) at 0.67\,GPa. Additionally the temperature derivative of the strain gauge resistance (black line, right scale) is shown, which is proportional to the thermal expansion coefficient $\alpha_c$ and our indicator for phase transitions. \cite{Aoki2009}}
\end{figure}
\begin{figure}[h!]
\includegraphics[width=16pc]{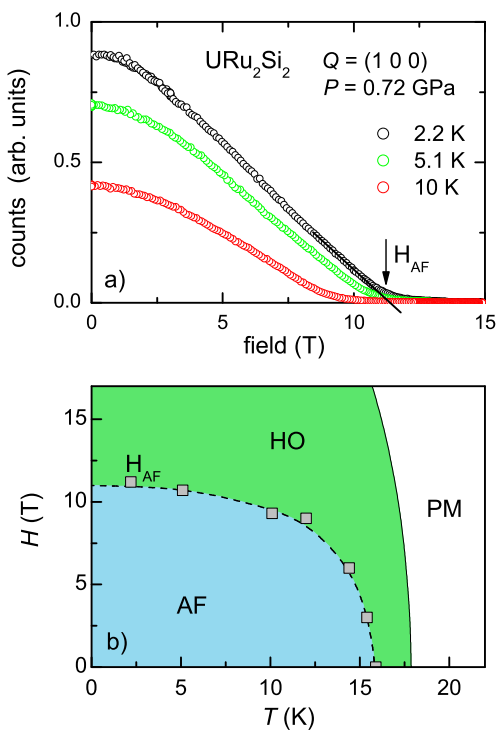}
\caption{\label{intensBandPDforArxive.eps}a) Field dependence of the intensity of the magnetic Bragg peak at $Q_0 = (1, 0, 0)$ at three temperatures at $P = 0.72$\,GPa. Above $H_{AF}$, the AF phase is completely destroyed. b) The phase diagram obtained from the curves on the left panel and from temperature sweeps at constant fields. The transition line between HO and PM follows the behavior measured by thermal expansion at 0.75\,GPa interpolated with high field data at 0.8\,GPa \cite{Villaume2008, Aoki2009,Jo2008}.}
\end{figure}
restricted to the field evolution of the $Q_0$ excitations. The reduction of signal by the combined use of a pressure cell in a cryomagnet device (high magnetic field up to 14.5\,T) led us to prepare a large crystal of 6\,mm diameter and 11\,mm length oriented with the easy c magnetization axis along the magnetic field H; the selected transmitted medium was a mixture of fluorinert FC 84/FC 87. As before, the selected pressure $P \approx 0.72$\,GPa was verified via the detection of the phase transitions PM-HO-AF via the thermal expansion measured by strain gauge. 
Figure \ref{n+t} shows the temperature dependence of the magnetic Bragg peak intensity (measurement of the intensity at the $Q_0 = (1, 0, 0)$ position) at zero field at 0.72\,GPa (black dots). By comparison, the signal obtained with the previous pressure experiments is drawn as well as the relative variation of the thermal expansion coefficient. It is obvious that pressure inhomogeneities exist in the present set up and that furthermore the deviation from hydrostatics is larger than the previous set up. From the $H$ and $T$ evolution of the Bragg reflection, we can estimate that 10\% of "parasitic HO"  phase remains even at $T = 0$ and $P = 0.72$\,GPa while ideal conditions will lead to the complete disappearence of the HO component as in the previous measurement (see figure \ref{Fig2}, left panel). However, following the field evolution of the intensity of the Bragg reflection at different temperatures (figure \ref{n+p}a), a complete disappearence of AF occurs above $H_{AF}$($T =$ 0) = 11\,T. The figure \ref{n+p}b represents the temperature evolution of the field $H_{AF}(T)$ where AF disappears.
\begin{figure}[h!]
\includegraphics[width=17pc]{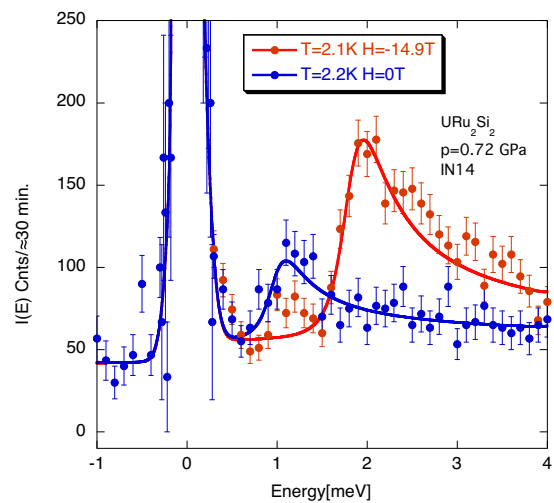}
\caption{\label{fig2}Energy spectra at the wave vector $Q_0 = (1, 0, 0)$ without magnetic field (blue dots) and in 14.9\,T (red dots) at 0.72\,GPa. The lines are guides to the eye.\cite{Aoki2009}}
\end{figure}
\begin{figure}[h!]
\includegraphics[width=17pc]{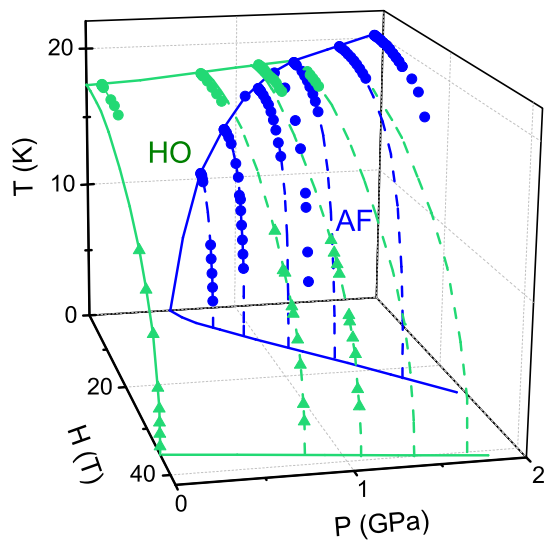}
\caption{\label{3D}Three dimensional phase diagram from thermal expansion measurements \cite{Aoki2009}. The high field data (green triangles) is from reference \cite{Jo2008}. The lines are guides to the eye.}
\end{figure}

Inelastic neutron scattering experiments lead to the interesting result that a nice excitation reemerges above $H_{AF}$ for $Q_0 = (1, 0, 0)$ (shown in figure \ref{fig2}), the weak contribution detected at $H = 0$ 
being a consequence of the 10 \% parasitic HO survivance. Taking into account our previous $P$ study, our proposal is that the disappearance of AF is associated with the field reentrance of the HO phase. Increasing further the magnetic field, the HO will be destroyed above $H_M \approx 35$\,T (the field to suppress the HO at zero pressure is $H_M \approx 35$\,T and increases with pressure \cite{Jo2008}). In this high field domain, the original higher carrier number of the PM phase will be recovered and different phases occur\cite{Kim2003}.

Neutron scattering experiments are crucial to point out the nature of the ($P$, $H$) induced phases, however it is very arduous to realize a systematic pressure study. This was achieved by detecting the ($P$,$H$) evolution of the transition temperatures by thermal expansion via a strain gauge as represented in figure \ref{3D}.

\begin{figure}[h!]
\includegraphics[width=17pc]{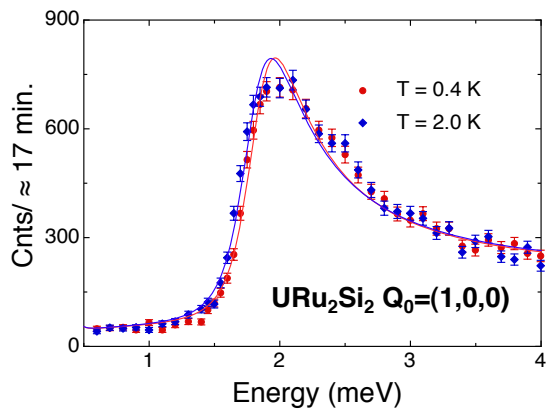}
\caption{\label{excitations}Energy spectra at the position $Q_0 = (1, 0, 0)$ above the superconducting transition at 2\,K and below it at 0.4\,K at ambient pressure. The lines are guides to the eye.}
\end{figure}

At $P_x$ bulk superconductivity collapses as well as the excitation at $Q_0 = (1, 0, 0)$. Surprisingly, no inelastic neutron scattering experiment at this wavevector has been 
reported in the SC phase. That led us to conduct an accurate inelastic scattering experiment with a high quality crystal ($T_{SC} \approx 1.2$\,K from the measurement of its diamagnetic shielding) down to 0.4\,K on IN12 spectrometer at ILL \cite{Bourdarottbp}. The temperature dependence of the inelastic spectrum for both $Q_0$ and $Q_1$ excitations were recorded. Figure \ref{excitations} shows the intensity of the inelastic response at $Q_0$ for $T = 2$\,K ($> T_{SC}$) and $T = 0.4$\,K ($< T_{SC}$). At 2\,K, no quasielastic contribution can be detected at
low energy; the same lack of quasielastic component was observed for the energy response at $Q_1$. At a first glance, no drastic change occurs below $T_{SC}$ at $T = 0.4$\,K for both wavevectors $Q_0$ and $Q_1$: no emergence of a low energy resonance characteristic of the superconducting pairing as observed for the heavy fermion superconductors UPd$_2$Al$_3$, CeCu$_2$Si$_2$ and CeCoIn$_5$ \cite{Sato2001,Hiess2006,Stock2008,Stockert2008}.
Looking more closely, the excitation at $Q_0$ shifts by $40$\,$\mu$eV, while the excitation at $Q_1$ remains unchanged. The shift in energy of the excitation at $Q_0$ through $T_{SC}$ as well as the collapse of the inelastic response above $P_x$ may suggest that the origin of the SC pairing can be magnetic excitations as proposed for UPd$_2$Al$_3$.

The absence of detectable,  low energy quasielastic ($T > T_{SC}$) or resonant ($T < T_{SC}$) contributions at Q$_0$ by comparison to UPd$_2$Al$_3$, CeCu$_2$Si$_2$ and CeCoIn$_5$ can be explained partly by the weakness of its Sommerfeld coefficient $\gamma$ ($\gamma = 70$\,mJmol$^{-1}$K$^{-2}$ in URu$_2$Si$_2$, 140\,mJmol$^{-1}$K$^{-2}$ in UPd$_2$Al$_3$, about 1000\,mJmol$^{-1}$K$^{-2}$ in CeCu$_2$Si$_2$ and CeCoIn$_5$\cite{Flouquet2005}). Indeed a weak $\gamma$ corresponds to a large relaxation rate for the normal state fluctuations ($T > T_{SC}$) and therefore the excitation spectrum is spread out in energy space. It is also possible that this magnetic response in \urs  is not strongly structured in momentum space as well contrarily to the three previous cases where the quasielastic response is concentrated along the antiferromagnetic hot spot. The case of \urs appears rather similar to that of the skutterudite system PrOs$_4$Sb$_{12}$ where despite a rather large $\gamma$-term ($\approx 300$\,mJmol$^{-1}$K$^{-2}$) no quasielastic contribution has been detected around any wavevector\cite{Kuwahara2005,Raymond2009}. For the case of PrOs$_4$Sb$_{12}$, it was recently stressed that the mass enhancement may originate from the aspherical Coulomb scattering of conductions electrons from the very low energy singlet-triplet crystalline electric field excitation \cite{Zwicknagl2009}. This mechanism is related to the dominant quadrupolar degrees of freedom at play in this compound.  Multipolar degreees of freedom may be involved in the formation of complex hidden order phase in \urs. 

On entering under pressure above $P_x$ in the AF phase, it was observed from resistivity measurements that the gap energy jumps to a higher value at $P_x$\cite{Aoki2009}. According to $P$ studies realized on 1-1-5 heavy fermion systems such as CeRhIn$_5$ on the duality between AF and SC, this jump can explain the collapse of SC in the AF phase\cite{Knebel2008}. Thus the proof that the $Q_0=(1, 0, 0)$ excitation may be the source of the SC pairing deserves more theoretical treatment.
\section{Conclusion}
New neutron scattering experiments show clearly that the excitation at $Q_0 = (1, 0, 0)$ is characteristic of the hidden order state. It disappears under pressure at $P_x$ where the ground state switches from HO to AF and is restored above $H_{AF}$ where the groundstate switches from AF to HO. There are strong indications from spin dynamics, but also from transport measurements that the wavevector of the HO phase is $Q_{AF} = (0, 0, 1)$. It is also demonstrated that the $Q_0$ excitation is directly coupled with superconductivity ($P$ and $T$ variation). Experimentally, \urs is a beautiful case where the main effect at the first order transition at $P_x$ occurs on the sublattice magnetization, on the inelastic neutron scattering response and on the thermal expansion, while other probes like the bulk magnetization, the resistivity, the Shubnikov-de Haas oscillations have only minor changes \cite{Nakashima2003}. For neutron scattering, the next targets are to determine the relation between the sharp excitations at $Q_0$ and $Q_1$ and the Fermi surface reconstruction. Theoretical work on the absence of quasielastic response, on the source of SC pairing and on the ($P$-$H$) instability of HO and AF phases is strongly needed.
\section*{Acknowledgements}{Financial support has been given by the French ANR within the programs ECCE and NEMSICOM.}
\providecommand{\newblock}{}

\end{document}